\documentstyle[aps,multicol,psfig]{revtex}

\begin{document}
\draft


\title{Comparisons of spectra determined using detector atoms and spatial correlation functions}
\author{M. Havukainen}

\address{
Helsinki Institute of Physics, PL 9, FIN-00014 Helsingin yliopisto, Finland
}


\date{\today}
\maketitle

\begin{abstract}

We show how two level atoms can be used to determine the local time
dependent spectrum. The method is applied to a one dimensional cavity.
The spectrum obtained is compared with the mode spectrum determined
using spatially filtered second order correlation functions. The
spectra obtained using two level atoms give identical results
with the mode spectrum. One benefit of the method is that only one time
averages are needed. It is also more closely related to a realistic
measurement scheme than any other definition of a time
dependent spectrum.

\end{abstract}




\begin{multicols}{2}
\narrowtext

\section{Introduction}
\label{Introduction}

It has turned out to be difficult to give a universally accepted
definition of the
time dependent spectrum. There are several definitions
\cite{page,lampard,silverman} which have not been popular in
optics. A definition, which is connected to a realistic
spectrum measurement was given by Eberly and W\'odkiewicz
\cite{eberly}. Their, 'physical spectrum', has become
a kind of canonical definition for a time dependent spectrum in
quantum optics. All definitions mentioned
above are based on integrals, typically Fourier transforms, over
different two time averages of classical stochastic variables or
quantum mechanical operators.

In our earlier paper \cite{analatompaper},
a totally new approach was introduced. Instead of calculating two time
averages the radiation is guided to a group of two level atoms.
All the atoms have equal, very small decay constants, but the resonance
frequencies are all different. The spectrum measured by the atoms
can be read from the excitation probabilities of the atoms.
We have studied
the spectrum of resonance fluorescence radiation from a laser
driven three level atom \cite{analatompaper}. The method was shown to
give exactly the same result as the 'physical spectrum' defined by
Eberly and
W\'odkiewicz. In order to calculate the two time averages needed to
determine the 'physical spectrum', the Quantum Regression Theorem (QRT)
\cite{lax}
must be used. It has been shown that there are systems where
QRT cannot be used i.e. the two time averages it gives are incorrect
\cite{oconnel}.
This is the case if the interaction between the matter, which emits
the radiation, and the radiation field is strong.
Interaction can be strong e.g. in microcavities. In the method of
analyzer atoms, two time averages are not needed.
The method can be used also in situations where the QRT is known to
give incorrect results. It can be said that it is closer to a realistic
spectrum measurement than any other definition.

In our earlier paper, the three level atom was coupled to the
analyzer atoms using cascaded master equations
\cite{cascadegardiner,cascadecarmichael}.
Thus the quantum mechanical
state of the field was not made available.
In this paper we connect the method of analyzer atoms to our
cavity QED simulations in one dimension. In these simulations
there are two level atoms inside a one dimensional cavity.
The field is quantized using canonical quantization. The
state vector is restricted to have only one excitation i.e.
the field strength is very low. The model was introduced
by V. Bu\v{z}ek et. al. \cite{buzekczech,buzekkorean} and has been used in
several different studies in one \cite{decay}
and two dimensions \cite{cavity2dpaper}.
The quantum mechanical state of the field is available so the
spectrum of the radiation can be read directly from the
excitations of the modes. The spatial dependence of the spectrum is
obtained by calculating the second order correlation function
$g(r_1,r_2)$ and using appropriate filter functions. The typical
situation in the simulations is that the initial field intensity
is split into left and right propagating parts using two level atoms.
The spectra of both parts is determined using both analyzer
atoms and filtered correlation functions. In all cases
the two different methods give identical results.

In Sec. II we specify the model used in the simulations. Spatial
correlation functions and their relation to the spectrum is
explained in Sec. III. In Sec. IV we present the analyzer atom
method for spectrum measurements. In Sec. IV we show several results of
our simulations and finally in Sec. VI we present our conclusions.

\section{Hamiltonians and State vectors}

\subsection{Hamiltonians}

In our simulations the mode functions in the one dimensional
cavity ($0\leq r\leq L$) are

\begin{equation}
\label{modefunctions}
G_n(r)=\sin(k_nr), \ \ k_n=n\frac{\pi}{L}, \ \ n=1,2,3...
\end{equation}
All mode functions are zero at the edges of the cavity, $r=0$ and $r=L$.
The electric
and magnetic field operators can be expanded using the mode functions
(\ref{modefunctions})

\begin{eqnarray}
\label{EandB}
\hat{E} & = & \sum\limits_{n=1}^{\infty}\left(\frac{\hbar\omega_n}{\epsilon_0L}\right)^{1/2}\sin(k_nr)(\hat{a}_n+\hat{a}_n^{\dagger}) \\
\hat{B} & = & i\sum\limits_{n=1}^{\infty}\left(\frac{\hbar\omega_n\mu_0}{L}\right)^{1/2}\sin(k_nr)(\hat{a}_n-\hat{a}_n^{\dagger}), \nonumber
\end{eqnarray}
where $\hat{a}$ and $\hat{a}^{\dagger}$ are annihilation and creation operators.
The operators satisfy the usual canonical commutation relations

\begin{eqnarray}
[\hat{a}_n,\hat{a}_m^{\dag}] & = & \delta_{nm}, \\
\left[\hat{a}_n,\hat{a}_m\right] & = & [\hat{a}_n^{\dag},\hat{a}_m^{\dag}]=0.
\end{eqnarray}

For the energy density operator we get using the expansions (\ref{EandB})

\begin{eqnarray}
\hat{H}(r) & = & \frac{1}{2}\epsilon_0\hat{E}(r)^2+\frac{1}{2\mu_0}\hat{B}(r)^2 \\
& = & \frac{2\hbar}{L}\sum\limits_{n=1}^{\infty}\sum\limits_{n'=1}^{\infty}\sqrt{\omega_n\omega_{n'}}\sin(k_nr)\sin(k_{n'}r)(\hat{a}_{n'}^{\dagger}\hat{a}_n+\frac{1}{2}).\nonumber
\end{eqnarray}
Integration of $\hat{H}(r)$ over the whole cavity and use of the
orthogonal integral

\begin{equation}
\label{orthogonalintegral}
\int\limits_0^L\sin(k_nr)\sin(k_{n'}r)dr=\frac{L}{2}\delta_{nm}, \ \ \ n,m\geq 1
\end{equation}
gives the familiar field Hamiltonian

\begin{equation}
\hat{H}_F=\sum\limits_{n=1}^{\infty}\hbar\omega_n(\hat{a}_n^{\dagger}\hat{a}_n + \frac{1}{2}).
\end{equation}

At fixed positions inside the cavity there are $N_A$ two level atoms
with resonance frequencies $\omega_j$ and dipole constants $D_j$.
The atomic Hamiltonian is the sum over all one atom Hamiltonians

\begin{equation}
\hat{H}_A=\sum\limits_{j=1}^{N_A}\hbar\omega_j\hat{\sigma}_z^j,
\end{equation}
where $\hat{\sigma_z}$ is a
Pauli spin matrix. The radiation field and the atoms are coupled through
dipole coupling. The dipole operator of the $j$:th atom is 

\begin{equation}
\label{dipoleoperator}
\hat{D}_j=(D_j\hat{\sigma}_+^j + D_j^*\hat{\sigma}_-^j).
\end{equation}
For the interaction Hamiltonian we get

\begin{equation}
\hat{H}_I=-\sum\limits_{j=1}^{N_A}\hat{D}_j\cdot\hat{E}(r_j),
\end{equation}
where $\hat{E}(r_j)$ is the electric field operator (\ref{EandB}) at the atomic
position $r_j$. Using the expansion (\ref{EandB}) for the field operator and
(\ref{dipoleoperator}) for the dipole operator we get

\begin{equation}
\label{jaynescummings}
\hat{H}_I=-\sum\limits_{j=1}^{N_A}\sum\limits_{n=1}^{\infty}\left(\frac{\hbar\omega_n}{\epsilon_0L}\right)^{1/2}\sin(k_nr)(D_j\hat{\sigma}_+^j\hat{a}_n + D_j^*\hat{\sigma}^j_-\hat{a}_n^{\dag}).
\end{equation}
The terms $\hat{\sigma}_+^j\hat{a}_j^{\dag}$ and
$\hat{\sigma}_-^j\hat{a}_j$, which do not affect the time evolution
significantly, have been neglected. This approximation is called the
Rotating Wave approximation (RWA). The Hamiltonian (\ref{jaynescummings})
has the familiar Jaynes-Cummings form.
The total Hamiltonian is the sum of field, atomic and interaction Hamiltonians:

\begin{equation}
\hat{H}=\hat{H}_F + \hat{H}_A + \hat{H}_I.
\end{equation}

\subsection{State vectors}

The general state vector is restricted to have only one excitation.
The most general state vector has the form

\begin{eqnarray}
\label{generalstatevector}
|\Psi\rangle & = & \sum\limits_{k}\left( c_{k}|1\rangle_{k}\prod\limits_{{k'}\neq {k}}|0\rangle_{k'}\right) \otimes\prod\limits_{j=1}^{N_A}|0\rangle_j\nonumber\\
& & +\prod\limits_{k}|0\rangle_{k}\otimes\sum\limits_{j=1}^{N_A}\left( c_j|1\rangle_j\prod\limits_{j'=1,j'\neq j}^{N_A}|0\rangle_{j'}\right) \\ 
& \equiv & \sum\limits_{k}c_{k}|1_{k},0\rangle + \sum\limits_{j=1}^{N_A}c_j|0,1_j\rangle.    \nonumber
\end{eqnarray}
In the first sum, one of the field basis vectors have one excitation and in the
second one, the excitation is in the atomic basis functions. If the initial state
vector
has only one excitation it does not obtain more excitations because the RWA was used
in the interaction Hamiltonian. The state vector at all times is given by
equation (\ref{generalstatevector}).

The state vector used as the initial state in our simulations is
of the form

\begin{equation}
\label{gaussianphoton}
|\Psi\rangle=\sum\limits_{k}(2\pi\sigma_k^2)^{(-1/4)}\exp\left(-ikr_0 - \frac{(k-k_0)^2}{4\sigma_k^2}\right)|1_k,0\rangle.
\end{equation}
The mode distribution $|c_k|^2$ is a Gaussian centered at $k=k_0$ with
a variance $\sigma_k^2$. The phase factor $e^{-ikr_0}$ is important.
It guarantees that the energy density distribution is also a Gaussian
centered at $r_0$.

\section{Correlation functions and the mode spectrum}

\subsection{Correlation functions}

The exact quantum mechanical state of the field is determined uniquely if all
correlation functions of the operators $\hat{E}(r)$ and $\hat{B}(r)$ are known.
In our simulations, because of
the special form of the state vector (\ref{generalstatevector}) the second
order correlation function
determines the field state uniquely. In the following we denote
$\epsilon_0=\mu_0=1$. Using the expansions (\ref{EandB}) we get for the
normally ordered correlation function

\end{multicols}
\widetext
\begin{eqnarray}
\label{correE}
\lefteqn{\langle\Psi(t)|:\hat{E}(r_1)\hat{E}(r_2):|\Psi(t)\rangle} \nonumber \\
& & \hspace{1.5cm}=\sum\limits_{n=1}^{\infty}\sum\limits_{m=1}^{\infty}\frac{1}{L}\sqrt{\omega_n\omega_m}\sin(k_nr_1)\sin(k_mr_2)\langle\Psi(t)|\hat{a}_n^{\dagger}\hat{a}_m + \hat{a}_m^{\dagger}\hat{a}_n|\Psi(t)\rangle \\
& & \hspace{1.5cm}=\sum\limits_{n=1}^{\infty}\sum\limits_{m=1}^{\infty}\frac{1}{L}\sqrt{\omega_n\omega_m}\sin(k_nr_1)\sin(k_mr_2)(c_n^*(t)c_m(t) + c_m^*(t)c_n(t)) \nonumber \\
& & \hspace{1.5cm}=T^*(r_1,t)T(r_2,t) + T(r_1,t)T^*(r_2,t) \nonumber
\end{eqnarray}

\begin{multicols}{2}
\narrowtext
where

\begin{equation}
\label{T}
T(r,t)=\sum\limits_{n=1}^{\infty}\sqrt{\frac{\omega_n}{L}}\sin(k_nr)c_n(t).
\end{equation}
A similar calculation gives

\begin{eqnarray}
\label{correB}
\lefteqn{\langle\Psi(t)|:\hat{B}(r_1)\hat{B}(r_2):|\Psi(t)\rangle} \\
& &  = T^*(r_1,t)T(r_2,t) - T^*(r_2,t)T(r_1,t).\nonumber
\end{eqnarray}
It is possible to invert the formula (\ref{T}) and get the mode coefficients
in Eq. (\ref{generalstatevector}). Multiplying both sides by
$\sin(k_mr)$ and using the orthogonality integral (\ref{orthogonalintegral}) we get

\begin{equation}
c_m(t)=\frac{2}{\sqrt{\omega_mL}}\int\limits_0^L\sin(k_mr)T(r,t)dr.
\end{equation}
So if $T(r,t)$ is known the mode coefficients can be calculated.
It is also possible to calculate the coefficients $c_n$ from the correlation
functions (\ref{correE}) and (\ref{correB}). Let us define

\begin{eqnarray}
\label{w}
\lefteqn{W(r_1,r_2)} \nonumber \\
& = &\langle\Psi(t)|\left( :\hat{E}(r_1)\hat{E}(r_2): + :\hat{B}(r_1)\hat{B}(r_2):\right) |\Psi(t)\rangle \nonumber\\
& = & T^*(r_1)T(r_2)
\end{eqnarray}
Integrating both sides by $\int\limits_o^L\sin(k_1r)\sin(k_pr)dr$ we get

\begin{equation}
\label{modecoefficientsfromw}
c_1^*c_p=\frac{4}{L\sqrt{\omega_1\omega_p}}\int\limits_0^Ldr_1\int\limits_0^Ldr_2\sin(k_1r_1)\sin(k_pr_2)W(r_1,r_2).
\end{equation}
This gives equations for each $c_p$. For $p=1$ we get $|c_1|^2$. We can choose
$c_1$ to be real and determine $c_1$. The rest of the coefficients are now
determined uniquely. Normalization can be used as a check of the calculation.

\subsection{A local mode spectrum}
\label{sectionlocalmodespectrum}

The mode spectrum $|c_k|^2$ gives the time
dependent spectrum of the radiation in the whole cavity. In order to get a spectrum
of radiation in some part of the cavity only we use filtered correlation
functions.
The correlation function (\ref{w}) is replaced by a filtered correlation function

\begin{equation}
\label{filteredw}
W_F(r_1,r_2;r_0) = g^F(r_1,r_2,r_0)W(r_1,r_2),
\end{equation}
where the filter function is a real-valued window function centered at
$r_0$. The method is similar to the Windowed Fourier Transform (WFT),
Ref. \cite{fundamentalsofwavelets,anintroductiontowavelets}, which is used to determine the
time dependent frequency distribution of a time dependent signal.
The mode spectrum is calculated using a filtered correlation
function (\ref{filteredw}) in equation (\ref{modecoefficientsfromw}).
The spectrum obtained is the mode spectrum of the radiation in
the part of the cavity where the filter function is nonzero.

A typical example of the filter function of one variable is a Gaussian filter

\begin{equation}
g^F(r;r_0)=(2\pi\sigma_r^2)^{-1/2}\exp\left(-\frac{(r-r_0)^2}{2\sigma_r^2}\right),
\end{equation}
where $\sigma_r^2$ is the variance of the window. Two dimensional
filter is a product of two one dimensional ones. Another filter used in our
simulations is a constant filter which is unity in some region and zero
everywhere else

\begin{equation}
r = \left\{ \begin{array}{ll}
             1\ \ \ \ \ \ \  & r_{min}\leq r\leq r_{max} \\
             0\ \ \ \ \ \ \  & r\leq r_{min},\ \ r\geq r_{max}
            \end{array}
     \right. .  
\end{equation}
The general behavior of the spectrum with all filter functions is such that the
broader the filter function, the smaller details can be seen in the
spectrum.

\section{Traditional definitions of the spectrum and analyzer atoms}

\subsection{Traditional definitions of the spectrum}
In the previous section the spectrum was calculated using '{\it spatial correlation
functions at a specific time value}'.
The traditional method to get the spectrum is to calculate the Fourier transform
of two time averages of field operators, i.e. use '{\it time correlation
functions at a specific point}'. For a stationary field the spectrum can be
defined as

\begin{equation}
\label{fourierspectrum}
S(\omega)=\int\limits_{-\infty}^{\infty}d\tau\langle\hat{E}(t)\hat{E}(t+\tau)\rangle e^{i\omega\tau}.
\end{equation}
This definition is valid only for stationary fields. There are many
generalizations
of a Fourier spectrum for nonstationary fields \cite{page,lampard,silverman}.
A method which takes also the measurement
scheme into account was presented by Eberly and W\'odkiewicz \cite{eberly}.
They define
a 'physical spectrum' to be a double Fourier transform of two time averages
multiplied by filter functions

\end{multicols}
\widetext

\begin{equation}
\label{physicalspectrum}
S_{phys}(t,\omega_f,\Gamma_f) = \Gamma_f^2\int\limits_0^{\infty}d\tau_1\int\limits_0^{\infty}d\tau_2\exp(-(\Gamma_f-i\omega_f)\tau_1)\exp(-(\Gamma_f+i\omega_f)\tau_2)\langle\hat{E}^*(t-\tau_1)\hat{E}(t-\tau_2)\rangle .
\end{equation}

\begin{multicols}{2}
\narrowtext
The frequency filters are placed in front of the photodetector. They allow only
the radiation with a certain frequency to pass to the detector. The
spectrum is obtained from the relative intensities measured by the detector
when the filters are tuned to different frequencies. The spectrum obtained
is time dependent and the definition incorporates the measurement scheme.
Our method to include time dependence to the spectrum is analogous to the method
shown in the previous section. The definition (\ref{physicalspectrum}) has
'{\it time filtered}' correlation functions whereas in the last section
'{\it spatially filtered}' correlation functions were used.

In many problems of
quantum optics the radiation field is traced out from the Hilbert space.
As a result the time evolution of the system is described by a master
equation, which does not have information about the field degrees of freedom.
Using this approach it is not possible to calculate the spatial correlation
functions; then the definitions with time correlation functions like
(\ref{fourierspectrum}) and (\ref{physicalspectrum}) are the only useful
definitions.

\subsection{Analyzer atoms}
\label{sectionanalyzeratomspectrum}
In our earlier paper \cite{analatompaper}, a totally different
approach to the time dependent spectrum was introduced.
The radiation from the system of interest is guided to a group of
$N$ two level atoms which all have the same very small decay constant $\Gamma$.
The resonance frequencies of the atoms are all different

\begin{equation}
\label{analatomfrequencies}
\omega_n=n\cdot\Delta\omega, \ \ \ \ \Delta\omega=\frac{\omega_{max}-\omega_{min}}{N-1}, \ \ \ \ n=1,2...N.
\end{equation}
Initially all atoms are in the ground state. The incoming radiation excites the
atoms. Because the decay constants are equal and very small, the excitation
probabilities of the
atoms are directly proportional to the intensity of the incoming radiation at
the resonance frequency. The normalized excitation as a function of resonance
frequency of the atoms gives the time dependent spectrum of the radiation.
A detailed description of the method and a comparison with a spectrum
(\ref{physicalspectrum}) calculated using two time averages can be
found in Ref. \cite{analatompaper}.
In these simulations cascaded master equations were used to describe
a three level atom and a group of analyzer atoms.
The two spectra were shown to give identical
results for the resonance fluorescence radiation of a three level atom.

The analyzer atom method can be used in our cavity simulations. Now the atoms
with frequencies given by Eq. (\ref{analatomfrequencies}) and very small decay
constants are located
at some point inside the cavity. As in the case of master equations, the
excitations should give the local time
dependent spectrum of the radiation which passes them. In the next section the
spectrum measured using detector atoms is compared with the mode spectrum
calculated using filtered spatial correlation functions.

\section{Simulations}

In this section we show results of spectrum measurements in three
different cases. In all simulations there is one atom or several
atoms at the center of the cavity. The initial field state is localized
on the left and is moving to the right. The atom at the center
splits the photon into left and right propagating parts. The
spectra of these two parts are measured separately using analyzer atoms
and filtered correlation functions. In the first simulation the initial
field state is the Gaussian (\ref{gaussianphoton}) and there is one
atom at the center. In the second simulation there are three atoms at
the center. In the last simulation there is only one center atom
and the field state is a superposition of several random Gaussian
states. The units in our simulations are chosen is such a way
that $c=\epsilon_0=\mu_0=\hbar=1$.

\subsection{A Gaussian photon and one atom}

\begin{figure}[htp]
\centerline{\psfig{file=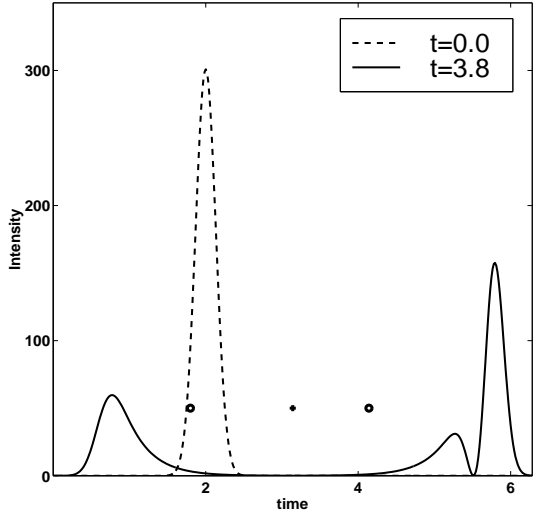,width=10.0cm,bbllx=1cm,bblly=1cm,bburx=21cm,bbury=27cm,angle=90,clip=}}
\caption{The energy density of the photon at $t=0.0$ and $t=3.8$. The initial
energy density is a Gaussian (\protect\ref{gaussianphoton}) with parameters
$r_0=2.0$, $k_0=100.0$ and $\Gamma_k=4\pi$. At $t=3.8$ there is one peak
propagating to the left and two to the right. The initial intensity is
split as a results of the interaction with the center atom, marked by a cross.
The circles show the positions of the analyzer atoms.
The length of the cavity is $L=2\pi$ and the number of
modes is $N_{mode}=400$.}
\label{intensityoneatom}
\end{figure}

\begin{figure}[htp]
\centerline{\psfig{file=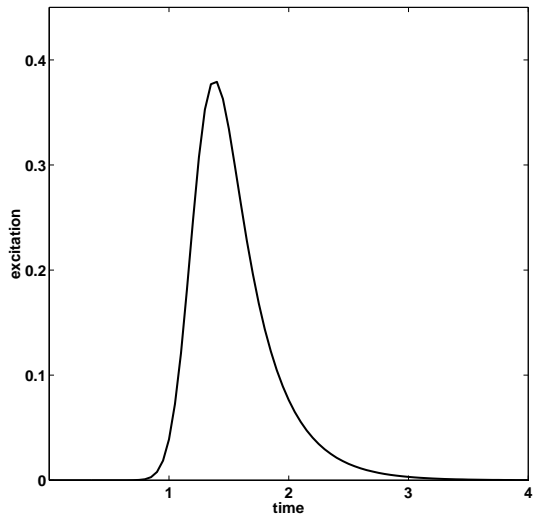,width=10.0cm,bbllx=1cm,bblly=1cm,bburx=21cm,bbury=27cm,angle=90,clip=}}
\caption{The excitation of the center atom. The photon excites the atom which
decays exponentially to the ground state. The decay constant of the atom is
$\Gamma=\pi$.}
\label{centeratomexcitation}
\end{figure}

In the first  simulation a Gaussian photon (\ref{gaussianphoton})
propagates towards
the center of the cavity. The length of the cavity is $L=2\pi$.
The mode distribution is a Gaussian
(\ref{gaussianphoton}), centered at
$k_0=100.0$ with the width $\Gamma_{ph}=4\pi$. The intensity
profile is centered at $r_0=2.0$. The initial intensity is shown in
Fig. \ref{intensityoneatom}.
\vfill
\pagebreak
\vspace{0.5cm}
In the center at $r=\frac{L}{2}$, there
is a two level
atom which is exactly on resonance with the photon $\omega_0=100.0$. The decay
constant of the atom is $\Gamma=\pi$, so the linewidth of the atom is
narrower than the width of the mode distribution of the field.
When the photon reaches the atom, the atom gets a nonzero excitation.
Later the atom emits energy back to the field modes via decay to the
ground state. The population of the excited state as a
function of time is shown in Fig. \ref{centeratomexcitation}.
\vspace{-0.5cm}
As a result of the interaction, part
of the energy is reflected to the left and part is able to pass the atom. The
intensity at $t=3.8$, after the interaction, is shown in
Fig. \ref{intensityoneatom}. The intensity
profile on the right has two peaks. The first peak has propagated to the right
without interaction and the second is the result of the atomic decay.

In addition to the center atom there are $600$ atoms which detect the spectrum
of the radiation which passes them as was explained in section
IV. At $r=1.8$
i.e. left from the center atom there are $200$ atoms which measure the
spectrum of the reflected radiation. The dipole couplings of the atoms are time
dependent in such a way that they are zero when the initial Gaussian photon
propagates to the right. At $t=1.5$ the atomic dipoles get the values
determined by (\ref{analatomfrequencies}) and start to detect the spectrum
of the radiation. On the right at $r=\frac{L}{2}+1.0$ there are two
groups of analyzer atoms.  The first group ($200$ atoms) measures the spectrum
all the time. After the first peak on the right has passed them, its
spectrum can be read from the excitations.
After the second peak has passed them the atoms give
the spectrum of all radiation right from the atom i.e. spectrum of the
first and second peaks. Another group at the
same position $r=L/2.0+1.0$ detects the spectrum of the second peak on the
right. The dipole constants of these atoms get nonzero values
(\ref{analatomfrequencies}) immediately after the first peak has
passed them.

\begin{figure}[htp]
\centerline{\psfig{file=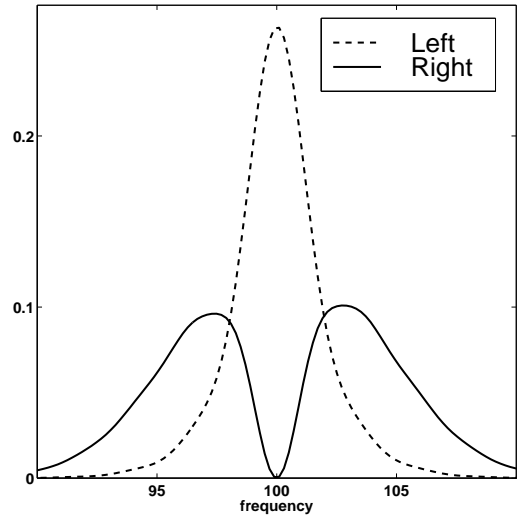,width=10.0cm,bbllx=1cm,bblly=1cm,bburx=21cm,bbury=27cm,angle=90,clip=}}
\caption{The normalized spectra of the radiation on the left and right measured
by the analyzer atoms,
Fig. \protect\ref{intensityoneatom}. The resonant radiation has been reflected
to the left and the off-resonant radiation has propagated to the right.
The spectrum has been read from the atoms after all radiation
has propagated through them. Both spectra have been normalized in such a way that
the area under the curves is unity.}
\label{analatomspecleftandrightoneatom}
\end{figure}

The spectra of the total intensity on the right and left measured using the
atoms are shown in Fig. \ref{analatomspecleftandrightoneatom}.
On the right the spectrum has a two peak
structure. It is interesting that on the right there is no intensity at the
resonance frequency of the atom. The center atom has been able to
reflect the resonant radiation to the left and only off-resonant radiation
passes the atom. The fact that resonant radiation is missing
is interesting because the second peak of the intensity
profile is the result of the atomic decay.

\begin{figure}[htp]
\centerline{\psfig{file=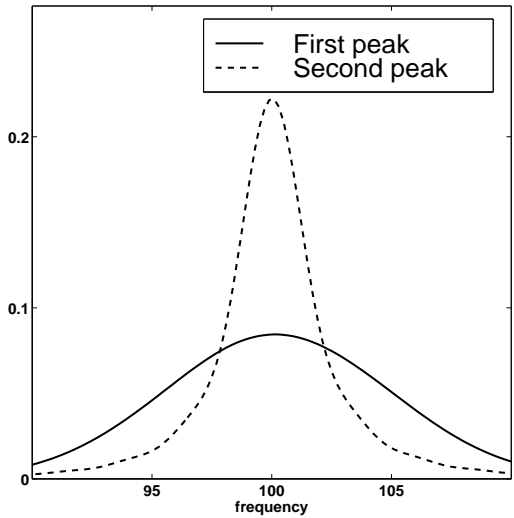,width=10.0cm,bbllx=1cm,bblly=1cm,bburx=21cm,bbury=27cm,angle=90,clip=}}
\caption{The measured spectra of the two peaks on the right (Fig.
\protect\ref{intensityoneatom}) separately. The spectrum of the first peak
is broad because the intensity profile is narrow. The second peak has
a typical Lorentzian spectrum of free decay. When the spectrum of both
peaks is measured, as in Fig. \protect\ref{analatomspecleftandrightoneatom},
the second peak creates an stimulated decay type of effect to the
detector atoms and the measured spectrum has a dip in the middle.
The normalization of the spectra is the same as in
Fig. \protect\ref{analatomspecleftandrightoneatom}}
\label{analatomspectwopeaksoneatom}
\end{figure}

As was explained earlier, we also measured the spectra of the two peaks on the
right separately. The result is shown in Fig. \ref{analatomspectwopeaksoneatom}.
The spectrum of the
first peak is quite close to the initial spectrum. The spectrum of the
second peak is narrower and is similar to the spectrum of the spontaneous
decay. The fact that the spectrum of the second peak is narrower is
understandable because the intensity profile is wider and the
interaction time with the detector atoms longer. The combined spectrum
of the two peaks has a dip in the middle, Fig.
\ref{analatomspecleftandrightoneatom}, whereas the spectra of the peaks
measured separately both have a single peak structure. This shows that
the spectrum is not additive. The second peak causes a stimulated
decay type of effect and the excitations of the detector atoms, which
are on resonance with the radiation, get smaller.

\begin{figure}[htp]
\centerline{\psfig{file=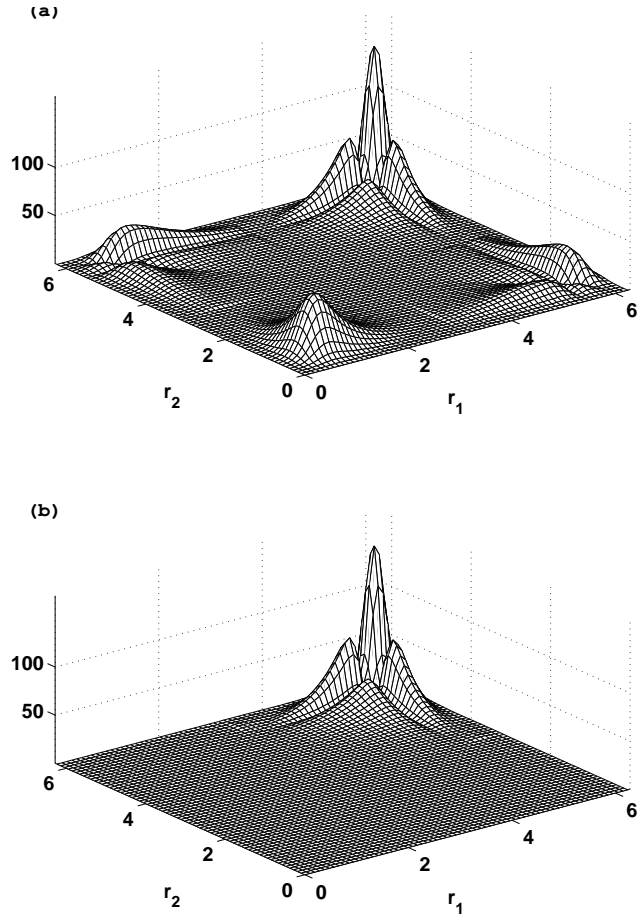,width=10.0cm,bbllx=1cm,bblly=1cm,bburx=21cm,bbury=27cm,angle=0,clip=}}
\caption{The correlation function $\langle\Psi|:\hat{E}(r_1)\hat{E}(r_2):|\Psi\rangle$
of the radiation at $t=3.8$ (a). The diagonal elements give the
energy density of the radiation, Fig. \protect\ref{intensityoneatom}. The
lower figure (b) shows the filtered correlation function used to determine the
mode spectrum on the right.}
\label{corre1and2}
\end{figure}

Next we calculate the mode spectrum using the filtered correlation
functions as was explained in Sec. III.
The normally ordered
correlation function $\langle:\hat{E}(r_1)\hat{E}(r_2):\rangle$ at
$t=3.8$ is shown in Fig. \ref{corre1and2} (a).
The diagonal elements $r_1=r_2$ give
the energy density of the radiation. Off-diagonal elements show
the coherence of the radiation. In order to calculate the spectrum
of the radiation in the left part of the cavity, a filtered correlation
function is used. The part $r_1, r_2 \geq \frac{L}{2}$ is replaced by
zero. The filtered correlation function is shown in Fig. \ref{corre1and2}
 (b).

Similarly in order to get the spectrum of the radiation on the right,
the replacement $\langle:\hat{E}(r_1)\hat{E}(r_2):\rangle=0$
when $r_1, r_2 \leq \frac{L}{2}$ is used.
The mode spectrum is calculated using the formula (\ref{modecoefficientsfromw})
with a filtered correlation function. The mode spectra obtained are
identical to the spectra measured using the analyzer atoms, Fig. 
\ref{analatomspecleftandrightoneatom}.
We calculated also the mode spectra of the two peaks on the right
separately using the appropriate filtered correlation functions.
Also these spectra were the same as measured by analyzer atoms, Fig. 
\ref{analatomspectwopeaksoneatom}.

\subsection{A Gaussian photon and three atoms}

\begin{figure}[htp]
\centerline{\psfig{file=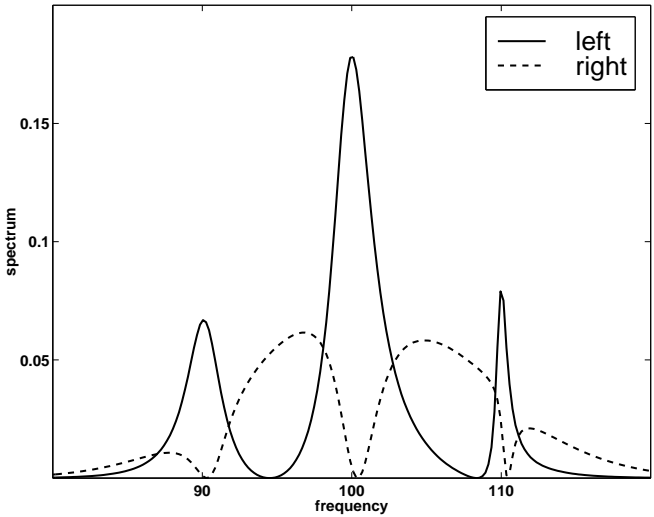,width=10.0cm,bbllx=1cm,bblly=1cm,bburx=21cm,bbury=27cm,angle=90,clip=}}
\caption{The measured spectra of the radiation on the left and right in the
simulation when there are three atoms at the center of the cavity.
The radiation on the left
has three peaks at resonance frequencies. On the right the broad spectrum
has three holes. The width of the peak (or hole) at $\omega_0=110.0$ is
narrower than the two others. The length of the cavity is $L=8\pi$ and the
number of modes $N_{mode}=1600$. The normalization of the spectra is
the same as in Fig. \protect\ref{analatomspecleftandrightoneatom}}
\label{analatomspecleftandrightthreeatoms}
\end{figure}

Next we add two more atoms to the center with resonance frequencies
and decay constants $\omega_0=90.0$, $\Gamma=\pi$ and $\omega_0=110.0$,
$\Gamma=\pi/4$ respectively.
The third atom is the same as earlier, $\omega_0=100.0$ and $\Gamma=\pi$.
The atom with resonance
frequency $\omega_0=110.0$ has a smaller decay constant than
the other two atoms. The width of the mode spectrum of the
Gaussian initial photon is increased to $\Gamma_{ph}=8\pi$. Because the
spectrum is very broad, the energy density profile of the photon
is narrow. We have carried out exactly the same spectrum measurement as in the
previous simulation for the radiation which passes the atoms and for the
reflected radiation. The results are shown in Fig.
\ref{analatomspecleftandrightthreeatoms}.
On the left the spectrum has three peaks at the resonance frequency. On the
right the original Gaussian spectrum has three holes at the atomic
frequencies. The width of the peak and the hole at $\omega_0=110$ is
narrower as was expected based on the atomic parameters.
The result is qualitatively the same as in the simulation with one
atom in the center. The resonant radiation is reflected and
the off-resonant radiation is able to pass the atoms. As in the previous
simulation we calculated the two spectra using also the filtered
correlation functions. The correlation function has a more complicated
form than the one in Fig. \ref{corre1and2}. The result is again identical
to the spectra given by the analyzer atoms.

\subsection{A random photon and one atom}

\begin{figure}[htp]
\centerline{\psfig{file=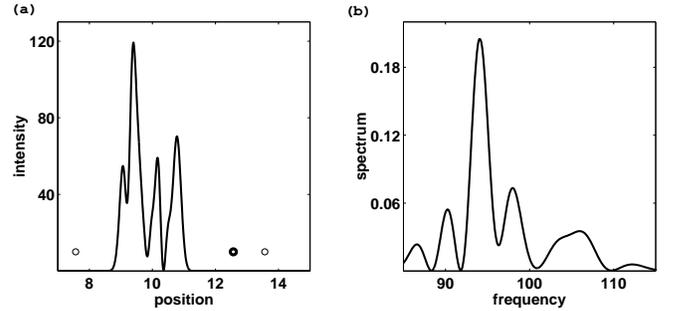,width=10cm,bbllx=1cm,bblly=1cm,bburx=21cm,bbury=27cm,angle=90,clip=}}
\caption{The intensity profile (a) and the mode spectrum (b) of the field
state which is a sum of ten Gaussian distributions for the photon.
The field is propagating
to the right. Note that only a part of the cavity is shown. 
At the center, $r=\frac{L}{2}$, there is one atom with the resonance
frequency $\omega_0=100.0$. The circles on the left and right are analyzer
atom which detect the spectrum. The initial mode spectrum, (b), has a 
six peak structure as some of the Gaussians are overlapping.
The normalization of the spectrum is the same as in
Fig. \protect\ref{analatomspecleftandrightoneatom}
}
\label{intensityandinitialmodespectrumstrange}
\end{figure}

In the two previous simulations the initial state of the photon
has been of the Gaussian form (\ref{gaussianphoton}). Next we
choose the initial state to be rather exotic.
The idea of this simulation is to test the method using a totally
different type of field excitation than in the previous simulations.
The initial photon is a superposition
of ten random Gaussian states. The parameters $k_0$, $\sigma^2_k$
and $r_0$ in equation (\ref{gaussianphoton}) are chosen randomly,
in such a way that the initial energy
density is on the left and propagating to the right. The frequency
distribution is centered at $\omega=100.0$.
Again at the center there is one atom which splits the intensity
into two parts. The initial intensity profile of the photon is shown in Fig.
\ref{intensityandinitialmodespectrumstrange}(a).
Note that the length of the cavity is now $L=8\pi$. The field has a four
peak structure and it propagates to the right towards the center
atom (thick circle). The atom has the resonance frequency $\omega_0=100.0$.
Circles right and left from the center atom show the positions
of the analyzer atoms which detect the spectrum. Figure
\ref{intensityandinitialmodespectrumstrange}(b)
shows the mode spectrum of the initial photon.

\begin{figure}[htp]
\centerline{\psfig{file=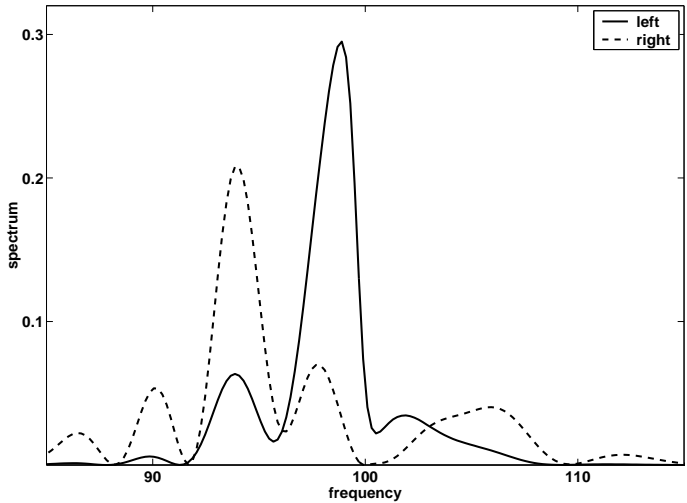,width=10cm,bbllx=1cm,bblly=1cm,bburx=21cm,bbury=27cm,angle=90,clip=}}
\caption{The spectra on the left and right measured using analyzer atoms.
The initial state of the field is a sum of ten random Gaussian states,
Fig. \protect\ref{intensityandinitialmodespectrumstrange}.
The spectrum
has been read from the atoms after all radiation has propagated through them.
The normalization of the spectrum is the same as in
Fig. \protect\ref{analatomspecleftandrightoneatom}}
\label{spectrumleftandrightstrange}
\end{figure}

As in the earlier simulations the center atom splits the radiation
into left and right propagating parts. The spectra of the two
parts are shown in Fig. \ref{spectrumleftandrightstrange}.

Because there is not much intensity which would be at resonance
with the center atom, the spectrum on the right is similar to
the original mode spectrum. Only a small fraction of the intensity
is reflected to the left. Both spectra have several peaks.
We calculated the mode spectrum on the left and right separately
using filter functions as earlier. Again the mode spectrum gives
exactly the same results as detected using analyzer atoms.

\section{Conclusion}
\label{Conclusion}

We have used two level atoms to detect the time
dependent and local spectrum of radiation in a one dimensional cavity.
The spectrum is read from the excitation probabilities of atoms with
very small decay constants. An
alternative method to determine the spectrum in this paper
has been to calculate
the mode coefficients using a filtered second order correlation
function. Different filter functions can be used to determine
the spectrum at a specific part of the cavity. It might be
possible to generalize the approach of filtered correlation
functions using wavelet expansions
\cite{fundamentalsofwavelets,anintroductiontowavelets,afriendlyguidetowavelets}.
In all cases studied, the spectra determined using these
two methods give identical results. In our earlier paper we
showed that the analyzer atom spectrum gives the same
result for a resonance fluorescence spectrum of a laser
driven three level atom as the 'physical spectrum' defined
by Eberly and W\'odkiewicz. This paper gives further proof
that the method really works.

One benefit of the method is that only one time averages
of quantum mechanical operators are needed. Thus it is closer
to a realistic spectrum measurement than the usual definitions
of time dependent spectrum, which typically require two
time averages. It can also be used in
situations were the usual method to calculate two time
averages in the Schr\"odinger picture, the quantum regression
theorem, is known to give incorrect results.

\section{Acknowledgements}

I want to thank the Academy of Finland for the financial support.
Computers of the Center for Scientific Computing (CSC) were used in the simulations.
Finally I want to thank V. Bu\v{z}ek, G. Drobn\'{y}, S. Stenholm and K.-A. Suominen for comments.

\end{multicols}

\end{document}